\documentclass[11pt,american]{article}

\usepackage[T1]{fontenc}
\usepackage[latin9]{inputenc}
\usepackage[a4paper]{geometry}
\usepackage{amsmath}
\usepackage{amssymb}
\usepackage{graphicx}
\usepackage{setspace}
\usepackage[authoryear]{natbib}
\makeatletter
\newcommand{\lyxaddress}[1]{
  \par {\raggedright #1 
  \vspace{1.4em}
  \noindent\par}
}

\usepackage[T1]{fontenc}

\makeatother

\usepackage{babel}
\begin{document}
\title{Estimating the energy requirements for long term memory formation}
\date{\today}
\author{Maxime Girard$^{1,3}$, Jiamu Jiang$^{2}$, Mark CW van Rossum$^{1,2}$}
\maketitle

\lyxaddress{1 School of Psychology\\
2 School of Mathematical Sciences\\
University of Nottingham, Nottingham NG7 2RD, United Kingdom\\
3 Integrative Biology and Physiology Master, Faculty of Sciences and
Engineering, Sorbonne University, Paris 75005, France}
\begin{abstract}
Brains consume metabolic energy to process information, but also to
store memories. The energy required for memory formation can be substantial,
for instance in fruit flies memory formation leads to a shorter lifespan
upon subsequent starvation \citep{Mery2005b}. Here we estimate that
the energy required corresponds to about 10mJ/bit and compare this
to biophysical estimates as well as energy requirements in computer
hardware. We conclude that while the reason behind it is not known,
biological memory storage is metabolically expensive, 
\end{abstract}
The human brain consumes some 20W of energy, 20\% of the body's total
consumption at rest. The cost for computation and information transmission,
mostly for synaptic transmission and spike generation, is well documented,
and the brain's design is now widely believed to be constrained by
energy needs \citep{Attwell:2001cs,Lennie2003,Harris2012,Karbowski2012}.
More recently the metabolic cost of learning has been added to the
brain's energy budget. Experiments in Drosophila indicate that these
costs are substantial. In \citet{Mery2005b} flies were exposed to
a classical conditioning protocol and learned to associate an odor
to a mechanical shock. After the protocol, all feeding was stopped
and the time to die from starvation was measured. It was found that
the conditioning reduced the lifespan compared to control flies. After
controlling for exposure to unconditioned and conditioned stimuli
separately, the decrease in lifespan was some 20\%.

Currently, it is not clear which neural processes are the main energy
consumers associated to learning and memory. However, it is known
that not all forms of memory are equally costly. Persistent forms,
such as Long Term Memory (LTM) in the fly, are costly, but the less
persistent Anaesthesia Resistant Memory (ARM) memory which decays
in a few days \citep{margulies2005deconstructing}, is not. Interestingly,
aversive LTM is halted under low energy conditions \citep{Placais13}.
Such adaptive regulation is also found in mammals where late-phase
Long Term Potentiation (late-LTP) is halted under low energy conditions,
while early phase LTP is not \citep{Potter2010}. 

In this note we review estimates for the energy required to store
a few bits of information, namely the association of an odour with
a noxious stimulus as happens in the protocol of \citet{Mery2005b}.
The estimates have large uncertainties that will hopefully be narrowed
down in the future. Nevertheless, we feel that these 'ball park' figures
are useful for theoretical considerations and future experiments.
We also discuss the estimate in the context of computer hardware.

How much information is stored in the classical odor-shock conditioning?
In order to learn the association of odor and shock requires at least
one bit of information, namely whether the stimulus is to be avoided
or not. If the valence of the stimuli were stored in more detail,
a few extra bits would be needed. Furthermore, the animal could store
the context of the stimulus, which would be functionally beneficial.
However, in contrast to mammals, we have not seen evidence for contextual
fear conditioning in flies. We therefore estimate that some 10 bits
are stored.

\subsection*{Direct measurement of energy intake after learning}

There are various methods to estimate the energy need for memory formation
from experiments. The first method is based on the fact that right
after learning, flies increase their sucrose intake to about double
the normal rate \citep[Fig1.c in][]{Placais17}. In the Capillary
Feeder assay (CAFE), the fly's energy uptake is determined from the
consumption of sugar water from a capillary (5\% sucrose; sucrose
carries 16.2kJ/g) \citep{rohtua}. The increase corresponds to an
additional intake of 42$\pm160$mJ (19$\pm85$mJ in Fig 6.e) compared
to control flies, where the errors denote standard deviations.\footnote{When comparing across experiments it should be noted that in the CAFE
assay, energy consumption is strongly reduced during an initial habituation
period during the first few days \citep{VandenBergh2022}.}

Of this energy intake, some will be lost due to metabolic inefficiency
and some will be lost in urine and feces. Assuming a 43\% conversion
efficiency to produce ATP \citep{PMID:32809434}, one can infer that
learning consumed some 20mJ in the form of ATP.

\subsection*{Estimation via lifetime}

An alternative estimate of the energy used for memory formation can
be found from the reduction in survival time upon starvation after
learning. It is simplest to assume that the fly dies whenever its
energy reserve $E(t)$ drops below zero. Next, assume that the energy
reserve decreases linearly in time with a rate $\beta$. I.e. $E(t)=E_{0}-\beta t$,
where $E_{0}$ is the initial energy reserve. Calorimetry can be used
to estimate the consumption rate for a non-starving fly at $\beta=(7\pm2)\mu$W
at 23 C. \citep{Fiorino2018}. (Noting that the metabolic rate varies
across strains and that the basal metabolic rate increases steeply
with increasing temperature; \citealp{Klepsatel2019}). The average
observed lifetime shortening, denoted $\Delta l$, caused by LTM memory
formation was about 4.5hrs in both the experiments of \citet{Mery2005b}
and \citet{Placais17}. Thus under this linear decrease model, one
finds $E_{\textrm{LTM}}=\beta\,\Delta l=(110\pm100)$mJ.

\begin{figure}
\centering{}\includegraphics[width=14cm]{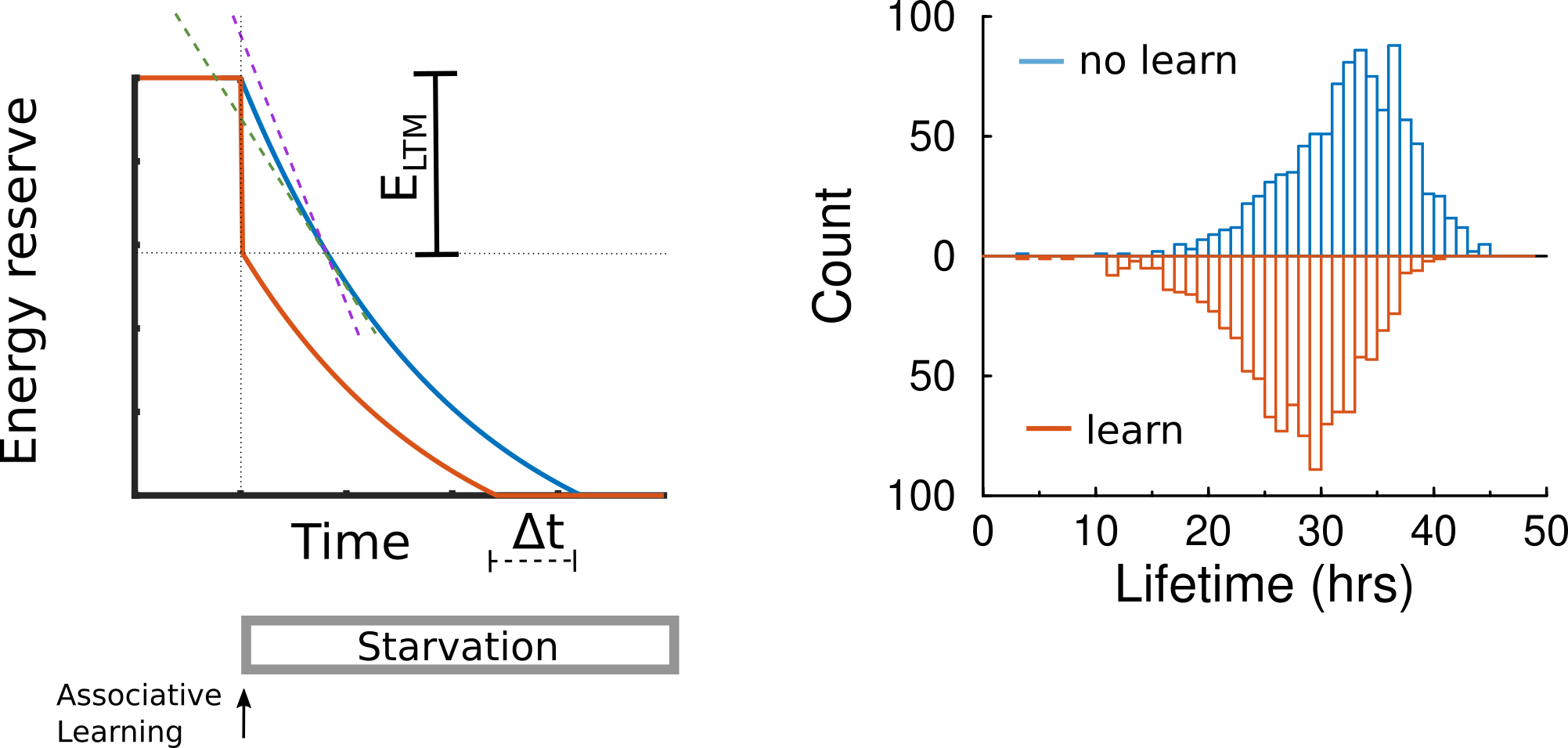}\caption{Left: Diagram for estimating the energy used for learning.\label{fig:Dia}
Energy reserve is plotted over time. At time 0, well fed flies either
learn an association, leading to a decrease in energy reserve (orange),
or are part of the control group (blue). Subsequently both group are
starved and die when their reserve hits zero; taught flies die an
interval $\Delta t$ earlier. The estimated energy used for learning
$E_{\textrm{LTM}}$ can be estimated from $\Delta t$ using either
the consumption rates right before (purple line) and right after learning
(green line). The true value falls between these estimates. Right:
Simulation of hazard model for 1000 flies in either the naive (top)
or learning population (bottom).}
\end{figure}

To examine the robustness of this estimate we add realistic features
to this model and show how this affects the estimate. First, the energy
consumption decreases as reserves are diminishing \citep{Fiorino2018}.
That is, the energy reserve is a convex function of time. Fig.~\ref{fig:Dia}
left shows the energy reserve versus time in two conditions. In the
control condition (blue curve) starvation start at time 0, causing
a gradual drop in the reserve. In the learning condition (orange curve)
learning causes a rapid drop in the reserve and takes place right
before starvation starts. 

We assume that metabolic rate is a function of the current energy
reserve only. This means that after expending energy on learning,
the energy reserve follows the same trajectory as that of a fly that
has been starved some time already. In other words, the learning associated
expenditure of the energy advances the energy trace by an amount $\Delta t$. 

We denote the initial rate of consumption as a positive number $\beta$
(slope of purple line; horizontally shifted horizontally for clarity),
and that after learning as rate $\beta'$ ($\beta'\leq\beta$; green
line). From Fig.~\ref{fig:Dia}, it can be seen that the energy estimate
is bounded as 
\[
\beta'\Delta t\leq E_{\textrm{LTM}}\leq\beta\Delta t
\]
This means that an estimate of the energy cost from lifetime differences
$\Delta t$ based on $\beta$, is possibly an over-estimate, but that
based on $\beta'$ is an underestimate. The metabolic rate after learning,
$\beta'$, has to our knowledge not been measured directly, however
in the setup of \citet{Fiorino2018} the metabolic rate drops some
30\% under a calorie restricted diet.

The calculation also holds when the energy consumption caused by learning
is not instantaneous as long as $\beta'$ is measured after the additional
consumption caused by learning has stopped.

\subsection*{Hazard model}

A more involved model to estimate the energy consumed by learning
is to use a hazard function formulation. A hazard function describes
the instantaneous probability for dying at a certain energy reserve
level \citep[see e.g.][]{Modarres1999,gerstner2002snm}. In the hazard
formulation, even if a population of flies all start with the same
energy reserve, they will die at different times. The most basic example
is a constant hazard. In that case the lifetimes are exponentially
distributed and the mean lifetime is the inverse of the hazard rate. 

We denote the hazard at a given energy reserve by $h(E)$. The hazard
increases as the energy reserve drops. We assume that the starvation
experiments are so drastic that any age dependence of the hazard can
be ignored. (Note that inclusion of age dependence would lead to further
underestimation of the energy -- in the extreme case that life time
is only age dependent, large changes in $E_{\textrm{LTM}}$ will not
affect lifespan).

In general, the mean lifetime $l$ in a hazard model is given by
\[
l=\int_{0}^{\infty}S(t)dt
\]
where the survival function $S(t)$ is given by $S(t)=\exp\left[-\int_{0}^{t}h(t')dt'\right]$.
We explore how advancing of the energy trace due to learning as in
Fig.\ref{fig:Dia} changes the average lifetime. With a tilde we denote
the quantities after learning. The advance means that $\tilde{h}(t)=h(t+\Delta t)$,
so that the survival function for the learning flies is
\begin{align*}
\tilde{S}(t) & =\exp\left[-\int_{0}^{t}h(t')dt'+\int_{0}^{\Delta t}h(t')dt'-\int_{t}^{t+\Delta t}h(t')dt'\right]
\end{align*}
For small $\Delta t$ this can be approximated as 
\begin{align*}
\tilde{S}(t) & \approx S(t)\left[1-\int_{0}^{\Delta t}h(t')dt'-\int_{t}^{t+\Delta t}h(t')dt'\right]\\
 & \approx S(t)[1-\Delta t.h(0)+\Delta t.h(t)]
\end{align*}
The average lifetime of the learned fly is $\tilde{l}=\int_{0}^{\infty}\tilde{S}(t)dt$.
Using that $\int_{0}^{\infty}S(t)h(t)dt=\int_{0}^{\infty}\frac{dS(t)}{dt}dt=1$,
the lifetime after spending an amount $E_{\textrm{LTM}}$ at time
zero is reduced to
\begin{equation}
\tilde{l}=l+\Delta t\left[1-h(0)\,l\right]\label{eq:haz_dl}
\end{equation}

It is instructive to study the two limiting cases: When the hazard
is independent of energy and hence constant in time ($h(t)=h(0)$),
one has $\tilde{l}=l$. In that case there is no change in lifetime.
In the other case, when there is no hazard before starvation, that
is, $h(0)=0$, one has $\tilde{l}=\Delta t$. In general the lifetime
change $\Delta l=l-\tilde{l}$ will range between $0$ and the shift
in the energy profile $\Delta t$. Combined with the above result,
\[
E_{\textrm{LTM}}\geq\beta'\Delta t\geq\beta'\Delta l
\]
Thus using consumption rate $\beta'$, one will always \emph{underestimate
}the energy expended on learning. The hazard formulation always exaggerates
the underestimate. With the caveat of the unknown rate $\beta'$,
we conclude that $E_{\textrm{LTM}}\gtrsim100mJ$.

As an illustration of this model we simulated 1000 flies, Fig.\ref{fig:Dia}b.
The initial reserve was set to 0.6J and we assumed it decayed exponentially
as $\gamma dE/dt=-E-c$, where $\gamma=40$hrs and $c=0.3$J. The
hazard was modeled as $h=\exp(-kE)$/hr, with $k=20J^{-1}$. This
resulted in a lifetime of 32.3 hrs without learning, and 27.6hrs when
learning. The estimated expenditure ($\beta'\Delta l$) was 95mJ,
compared to a true value of $E_{\textrm{LTM}}=$100mJ used in the
simulation.

\section*{Discussion}

In summary, we used two ways of estimating the amount of energy needed
to learn a simple association from behavioural data, namely from excess
sucrose consumption and from change in lifespan. Encouragingly, the
estimates yield comparable numbers on the order of 100mJ, or some
10mJ/bit. It is interesting to compare these costs to memory costs
in digital computers. Both data storage and data transmission from
CPU to memory cost substantial amounts of energy. In typical personal
computers the slowest, most persistent, and most energy costly storage
is farthest removed from the processor \citep{das2015slip}. For instance,
a typical modern Solid State Drive (SSD) can write up to 3GByte/s,
taking about 10W (Samsung 970). Hence the energy cost of storage on
an SSD is about 0.5 nJ/bit. A hierarchy of smaller and faster caches
(L3, L2, L1) speeds up read and write access of data that is repeatedly
used by the CPU. Energy costs of these are only of the order of pJ/bit
\citep{molka2010characterizing,das2015slip}. With the caveat is that
computers are highly optimized for processing large chunks of data,
memory storage in computers is therefore some 6$\ldots$7 orders of
magnitude less costly than biological memory storage.

Why is biological learning so metabolically demanding? Currently it
is not clear whether most energy is consumed on the synaptic level,
network level, or organism level. The biophysical cost of synaptic
plasticity in mammals was estimated by \citet{karbowski2019metabolic}.
The leading cost there is by far protein phosphorylation, which far
outweighs estimates for protein synthesis, transport costs and other
costs such as actin thread milling. It is estimated as $3\times10^{6}$ATP/synapse/min.
Hence the cost of increased phosphorylation in a single synapse during
1 hour would come to 9pJ. Even with 1000 synapses undergoing plasticity
this number is still 3 orders of magnitude below the behaviour based
estimates above. Moreover, phosphorylation is more characteristic
of early, inexpensive early phase LTP than of the expensive late phase
LTP. Interestingly, there is recent evidence for different metabolic
pathways for different types of plasticity \citep{Dembitskaya2022}.
So while those estimates are thus not inconsistent with our estimates,
a large amount of energy use remains unaccounted for. 

To determine if the missing energy is used directly by synaptic plasticity,
it would be of interest to measure energy consumption when the number
of modified synapses or the number of memoranda is varied. If learning
two associations would costs double the energy, synaptic processes
are likely the main consumer. In that case the energy needed to learn
multiple associations could rapidly become enormous. For instance
learning the well-known MNIST data set requires at some $10^{8}$
synaptic updates (in preparation). Saving strategies will be needed
in that case \citep{li2020energy}. 

An alternative is that the major consumers are changes in brain activity
by coordinated processes such as replay -- which contributes to memory
consolidation in mammals, but also in flies \citep{cognigni2018right}.
Calorimetry during learning could provide insight into such hypotheses.
Finally, behavioral or physiological changes resulting from the learning
protocol might explain the increased consumption. Control experiments
with unpaired stimuli in \citet{Mery2005b} might not have completely
corrected for such effects.

No matter the answer, animals are likely constrained by the high metabolic
cost of learning; their savings strategies will help to understand
biological memory formation. 

\subsubsection*{Acknowledgments}

We would like to thank Pjotr Dudek and William Levy for discussion.
Jiamu Jiang is supported by a Vice-Chancellor International award
from the University of Nottingham.

\bibliographystyle{plainnat_initialsonly}
\bibliography{/home/vrossum/neuro_jab}

\end{document}